\documentclass[10pt,journal,compsoc]{IEEEtran}
\usepackage{bm}
\usepackage{amssymb}
\usepackage{amsmath}
\usepackage{hyperref}
\usepackage{float}
\usepackage{booktabs}
\usepackage{multirow}
\usepackage{graphicx}
\usepackage[figurename=Figure]{caption}

\ifCLASSOPTIONcompsoc
  \usepackage[nocompress]{cite}
\else
  \usepackage{cite}
\fi
\ifCLASSINFOpdf
\else
\fi
\begin{document}
%\linenumbers
\title{A robust and generalizable immune-related signature for sepsis diagnostics}

\author{Yueran Yang$^{1,2,\#}$, Yu Zhang$^{1,2,\#}$, Shuai Li$^2$, Xubin Zheng$^{1,3}$, Man-Hon Wong$^3$, Kwong-Sak Leung$^3$, Lixin Cheng$^{1,*}$% <-this % stops a space
\IEEEcompsocitemizethanks{\IEEEcompsocthanksitem[1] Shenzhen People's Hospital, First Affiliated Hospital of Southern University of Science and Technology, The Second Clinical Medicine College of Jinan University, Shenzhen, China   
\IEEEcompsocthanksitem[2] John Hopcroft Center for Computer Science, Shanghai Jiao Tong University, Shanghai, China
\IEEEcompsocthanksitem[3] Department of Computer Science and Engineering, The Chinese University of Hong Kong, Shatin, New Territories, Hong Kong, China
\IEEEcompsocthanksitem[\#] These authors contribute equally to this study.
\IEEEcompsocthanksitem[*] Corresponding author: Lixin Cheng, Ph.D., Shenzhen People's Hospital, 1017 Dongmen North Road, Shenzhen, Guangdong Province 518020, China; Email: easonlcheng@gmail.com } }% <-this % stops an unwanted space


\markboth{IEEE\/ACM Transactions on Computational Biology and Bioinformatics}%
{}



\IEEEtitleabstractindextext{%
\begin{abstract}
  High-throughput sequencing can detect tens of thousands of genes in parallel, providing opportunities for improving the diagnostic accuracy of multiple diseases including sepsis, which is an aggressive inflammatory response to infection that can cause organ failure and death. Early screening of sepsis is essential in clinic, but no effective diagnostic biomarkers are available yet. Here, we present a novel method, Recurrent Logistic Regression, to identify diagnostic biomarkers for sepsis from the blood transcriptome data. A panel including five immune-related genes, LRRN3, IL2RB, FCER1A, TLR5, and S100A12, are determined as diagnostic biomarkers (LIFTS) for sepsis. LIFTS discriminates patients with sepsis from normal controls in high accuracy (AUROC = 0.9959 on average; IC = [0.9722-1.0]) on nine validation cohorts across three independent platforms, which outperforms existing markers. Our analysis determined an accurate prediction model and reproducible transcriptome biomarkers that can lay a foundation for clinical diagnostic tests and biological mechanistic studies.
\end{abstract}

}

\maketitle

\IEEEdisplaynontitleabstractindextext

% \fi
\IEEEpeerreviewmaketitle

\IEEEraisesectionheading{\section{Introduction}\label{sec:introduction}}

\IEEEPARstart{S}{epsis} is a life-threatening organ dysfunction caused by a host’s unbalanced response to an infection. It is one of the most severe diseases in the intensive care unit (ICU) and one of the world’s leading lethal diseases \cite{r1}\cite{r2}\cite{r3}\cite{Zheng_2021}. Its common clinical manifestations include abnormalities in body temperature, heart rate, breathing, and peripheral white blood cell counts. Besides, sepsis is often accompanied by multiple organ dysfunction syndromes, such as hemodynamic instability, respiratory failure, and disseminated intravascular coagulation. In the past few decades, the high morbidity and mortality caused by sepsis have made the society to endure huge economic burden \cite{r1}\cite{r2}\cite{r3}. The prevalent methods of the diagnosis of sepsis are microbiological culture and taxonomic identification of the pathogen. However, the methods based on bacterial culture techniques have several shortcomings: (1) it usually takes 24 hours to obtain a positive result; (2) only one-third of the blood cultures are positive in clinically diagnosed sepsis cases, so negative results in culture do not mean negative cases; (3) the chances of a positive culture are reduced in patients who have already used antibiotics; (4) false positives are frequently caused by contamination; (5) short-term bacteremia can lead to a positive blood culture without a severe inflammatory response. Therefore, the sensitivity and specificity of the methods based on microbiological culture are quite low, and hence fails to diagnose sepsis effectively.\cite{culture1}\cite{culture2}

Biomarkers such as procalcitonin (PCT) and C-reactive protein (CRP) have been considered to diagnose and evaluate sepsis in emergency department (ED) and intensive care unit (ICU). PCT is increasingly recognized as an important diagnostic and monitoring tool in clinical practice that provides significant information for clinical decision making. It is a potential biomarker in assisting clinicians in the diagnosis of generalized infection and sepsis in ICU. Several systematic reviews and meta-analyses have been carried out to describe the utility of PCT in distinguishing sepsis from SIRS and non-septic burn patients.\cite{marker1} However, the overall sensitivity and specificity of PCT range from 0.72 to 0.93 and 0.64 to 0.84, respectively \cite{Pierrakos_2020}, which is incompetent in the clinical context. CRP was reported as an indicator whose daily measurement is useful in monitoring sepsis, but its low specificity may be its primary drawback and it is hard to define CRP as an independent predictor of sepsis\cite{maker2}.

In recent years, with the rise of high-throughput sequencing technology, tens of thousands of genes can be detected in parallel \cite{r4}\cite{r5}\cite{r6}\cite{Nan_2020}, providing opportunities for precise molecular diagnosis using machine learning methods \cite{r7}\cite{r8}\cite{10.3389/fcell.2021.671302}\cite{Liu_2020}\cite{Wang_2019}. Several gene markers have been developed for the diagnostic prediction of sepsis. 
For instance, Scicluna \textit{et al.} proposed the FAIM3:PLAC8 ratio as a candidate biomarker to assist in the rapid diagnosis of community-acquired pneumonia (CAP) \cite{r9}, which accounts for a high proportion of intensive care unit (ICU) admissions for respiratory failure and sepsis. McHugh \textit{et al.} designed a classifier SeptiCyte Lab composed of four mRNAs of CEACAM4, LAMP1, PLA2G7, and PLAC8 by applying Support Vector Machine (SVM) and Random Forest \cite{r10}. Scicluna \textit{et al.} developed a sNIP score for sepsis diagnosis based on the expression abundance of three genes using SVM and bootstrapping \cite{r11}. However, the above-mentioned mRNA biomarkers cannot obtain consistent results in multiple independent data sets.

In this paper, we introduced a novel recurrent logistic regression (RLR) as an automatic detection for the diagnostic biomarkers of sepsis. Since patients with sepsis have a severely dysregulated immune system \cite{r10}\cite{r11}\cite{r12}, we principally concentrated on the immune-related genes (IRGs) and regarded them as the key molecular events involved in sepsis. Based on IRGs, the RLR model was trained and the less significant genes were filtered during each iteration until no gene is eliminated. Regularization and elimination of insignificant features were applied simultaneously in RLR to avoid overfitting and hence reduce the complexity of the discriminative model. The biomarkers identified by RLR were verified on nine independent expression cohorts across three different detection platforms. We also evaluated the classification performance of each individual gene in the identified biomarkers. Finally, network and functional analyses were carried out for the genes interacting with these biomarkers.

\section{Materials and Methods}
\subsection{Data and preprocessing}
Eleven different gene expression cohorts were collected from the Gene Expression Omnibus (GEO) database with both sepsis samples and healthy controls, including three adult datasets and eight pediatric datasets (Table \ref{tab:1}). In total, 1,384 samples were analyzed from three microarray platforms, Affymetrix Human Genome U133 Plus 2.0 (AffyU133P2), Affymetrix Human Genome U219 (AffyU219), and Agilent Human Gene Expression 4x44K v2 Microarray (AgilentV2). The raw data were preprocessed and normalized using the robust multichip average (RMA) algorithm \cite{r13}\cite{r14}\cite{r15}.

GSE57065 is adopted for model training (Discovery cohort \uppercase\expandafter{\romannumeral1}) and GSE26378 is used for tuning the hyperparameter (Discovery cohort \uppercase\expandafter{\romannumeral2}). Seven datasets (GSE95233, GSE28750, GSE8121, GSE13904, GSE26440, GSE9692, and GSE4607) from AffyU133P2 serve as the validation cohorts \uppercase\expandafter{\romannumeral1} to evaluate the diagnostic performance. GSE65682 and E-MTAB-
1548 detected by other platforms are set as the validation cohorts \uppercase\expandafter{\romannumeral2} for evaluating the cross-platform capability.

We intend to train a robust prediction model across the biological heterogeneity of childhood and adult sepsis, so the adult and children samples were incorporated for model training.

\subsection{Immune-related gene selection}
Since sepsis is a disease related to patients’ immune systems, only immune-related genes (IRGs) are considered as potential biomarkers in this study. 770 IRGs were collected from the database nanoString (www.nanoString.com), which has been used in hundreds of studies of pathogen infection and the related host response.\cite{nano}\cite{nano2} The numbers of IRGs are 737, 740, and 627 on AffyU133P2, AffyU219, and AgilentV2, respectively. We aimed to find a biomarker that can be applied to different platforms, so the 608 common IRGs of the three platforms were utilized for computational modeling (Figure \ref{fig:1}a).

\subsection{Recurrent logistic regression}
\begin{figure*}[h]
	\centering
	\includegraphics[width=\textwidth]{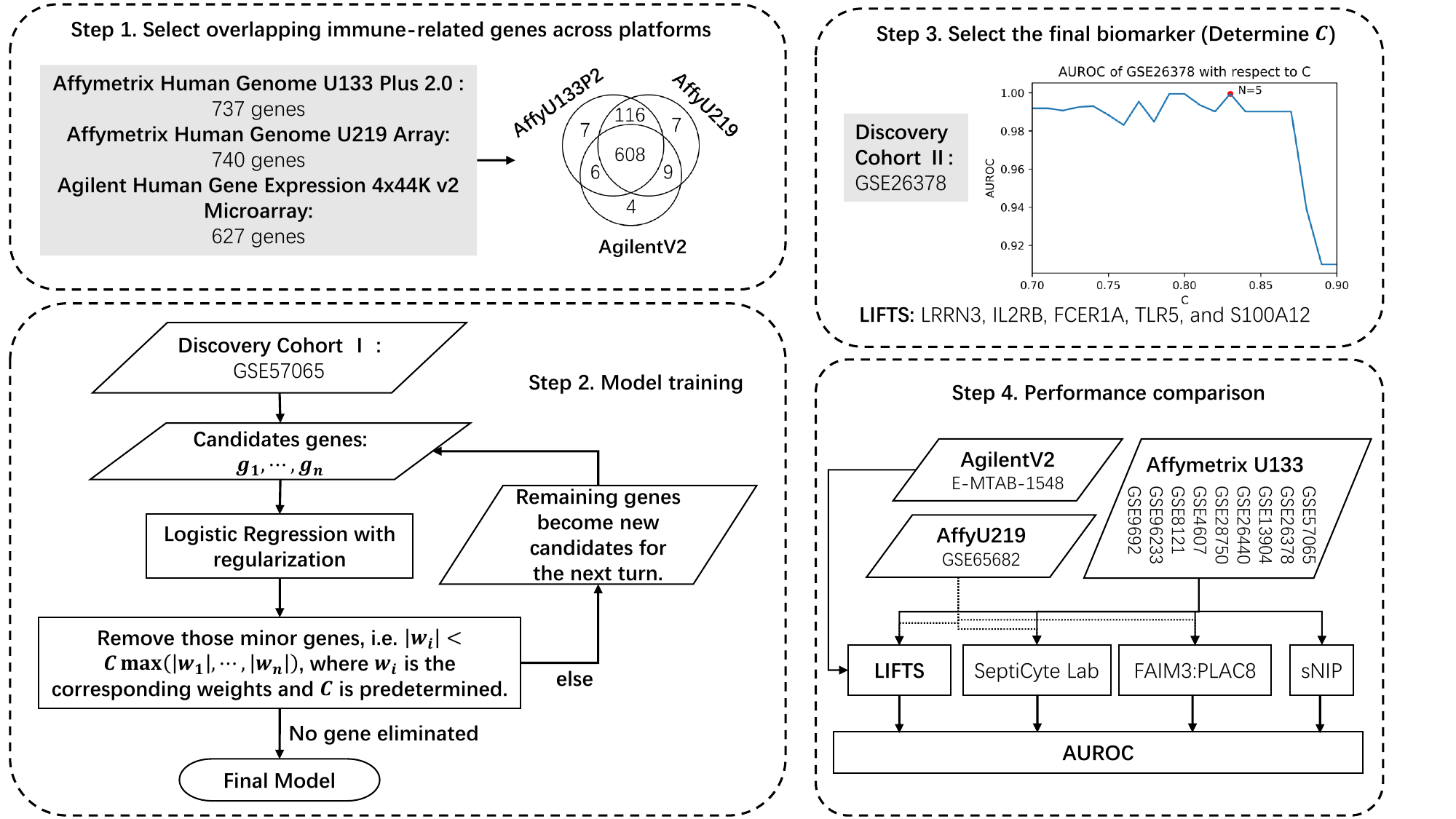}
	\caption{Workflow for the identification of sepsis biomarkers. a) statistics of immune-related genes of all cohorts in three platforms. b) the flow chart of the recurrent logistic regression algorithm including the regression step and the elimination step. c) determining the hyperparameter c with another discovery cohort and finalizing the biomarkers. d) validating and comparing diagnostic capability with distinct cohorts and platforms.}
	\label{fig:1}
\end{figure*}

Recurrent logistic regression contains many iterations with model optimization and automatical feature selection, since each iteration involves regression step and elimination step (Figure \ref{fig:1}b). 

\textbf{Regression step:} Logistic regression is employed to the candidate gene set (initially 608 IRGs). The expression abundance of genes in each sample, is represented by a vector denoted as
\begin{equation}
  \bm{x} = (g_1,\cdots,g_n)^T
\end{equation}
where $g_i$ is the $i$-th gene expression. 

To construct a classifier involving fewer genes features based on the expression vector $\bm{x}$ of a sample, a function $f:\mathbb{R}^n\to\{0,1\}$ is built, where $0$ represents healthy controls and $1$ represents sepsis. The logistic model applied in RLR is a binary classifier expressed by
\begin{equation}
  f(x)=\frac{1}{1+e^{-\bm{wx}}}=\frac{1}{1+e^{-(w_0+w_1g_1+\cdots+w_ng_n)}}
\end{equation}
where $\bm{x}$ is an expression vector and $f(x)$ is a diagnostic risk score used to predict the probability of having sepsis. $\bm{w} = (w_0,w_1,\cdots,w_n)^T$ are parameters optimized by the cost function with regularization,
\begin{equation}
  J(w;X)=\frac{1}{2} \bm{w}^T \bm{w}+\sum_{i=1}^n \log\left(1+\exp\left(-y_i(\bm{x}_i^T\bm{w}+w_0)\right)\right)
\end{equation}
where $X$ is the collection of samples $\bm{x}_1,\bm{x}_2,\cdots, \bm{x}_n$ in discovery cohort and $y\in\{0,1\}$ is the label for each sample.

\textbf{Elimination step:} After optimizing the regression model, the minor genes regarded as less significant are eliminated. A gene $g_i$ is defined as minor gene if the absolute value of its corresponding weight $w_i$ is less than a proportion of the absolute maximum weight, i.e.,
\begin{equation}
	|w_i|< C\max\left(|w_1|,\cdots, |w_n| \right)
\end{equation}
where $C\in [0,1]$ is a hyperparameter. Instead of using the traditional way that chooses a fixed threshold such as P value $< 0.01$, this step selects features adaptively based on the maximum weight trained by the model. 

The regression step and elimination step are repeated iteratively until it converges, specifically, no more minor gene remained. In this sense, the algorithm is named the recurrent logistic regression (RLR).

The RLR is first trained on the discovery cohort \uppercase\expandafter{\romannumeral1}  and evaluated by the AUROC on discovery cohort  \uppercase\expandafter{\romannumeral2}  (Figure \ref{fig:1}c). We exhaustedly tried possible values of the hyperparameter C with the search space between 0.75 and 0.9 and each interval equaling to 0.01. The hyperparameter C can therefore be determined by the optimal performance on discovery cohort  \uppercase\expandafter{\romannumeral2}.

\subsection{Performance evaluation and analysis}
Receiver operating characteristic (ROC) curve was applied for performance evaluation, which is the function image of True Positive Rate (TPR) with respect to False Positive Rate (FPR), where TPR represents the positive correctly classified samples to the total number of positive samples and FPR represents the ratio between the incorrectly classified negative samples to the total number of negative samples. Area Under the Curve (AUC) means the area under the ROC curve ranging from 0 to 1. Higher AUC indicates a more discriminative model. We use AUC to quantify the discrimination ability of the models on seven cohorts measured with the same platform and compare to the existing biomarkers. Moreover, the cross-platform capability is also evaluated on two cohorts from different platforms.

\begin{figure*}[h]
	\centering
	\includegraphics[width=\textwidth]{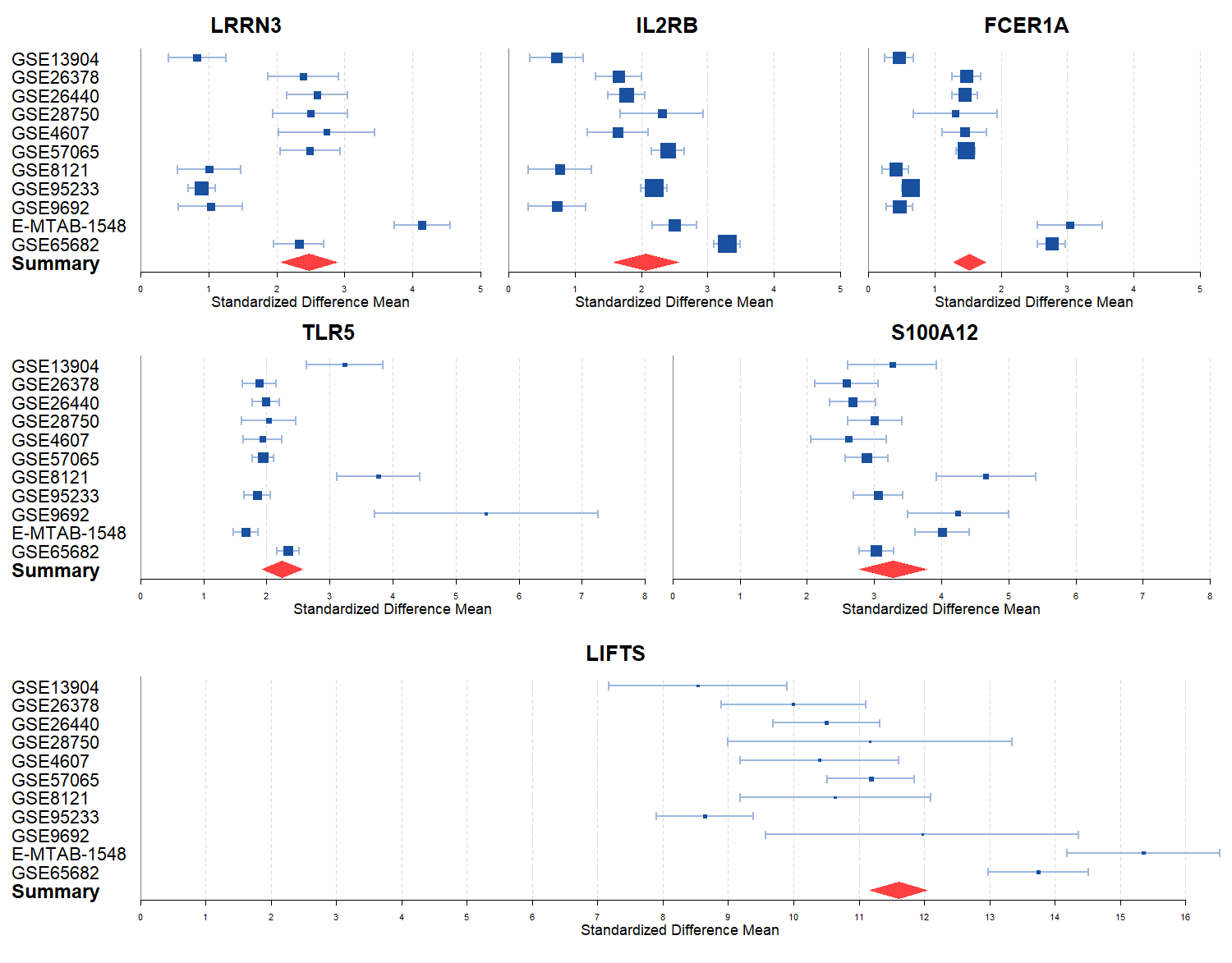}
	\caption{Forest plots of LIFTS and each genes in LIFTS. The x-axis represents the standardized mean difference between sepsis patients and healthy controls. The blue square is the average value of the difference and the size corresponds to the concentration of the data. The blue line represents the 95\% confidence interval. The red diamond represents the average difference of a given gene or LIFTS for all cohorts. The width of the diamond represents the 95\% confidence interval of the overall mean difference.}
	\label{fig:3}
\end{figure*}

Meta-analysis was conducted for the constructed gene panel LIFTS (LRRN3, IL2RB, FCER1A, TLR5, and S100A12) and the Standardized Mean Difference (SMD) are demonstrated in forest plot (Figure \ref{fig:3}). Four graphical elements are presented including the estimated SMD (solid block), the respective 95\% confidence intervals for each cohort (horizontal line), the non-effect size (vertical line), and overall estimation of all cohorts (red rhombus) \cite{r16}. 

To analyze the role of the five genes in LFTS, we presented the human protein interaction and constructed the protein interaction network. Protein interactions were obtained from the database InWeb\_InBioMap \cite{r17,r18}which is the most comprehensive resource for human protein interactome. Around 57\% of the interactions are experimentally validated. The interaction network was conducted and illustrated using the R package igraph.Function enrichment was carried out using the R package clusterprofiler \cite{r19} and the network-guided gene set characterization pipeline of KownEnG \cite{Blatti_2020}, respectively.

\section{Results}
\subsection{Identification of LIFTS}
Patients with sepsis have a severely dysregulated immune system, which impairs clearance of the infection and leaves the body susceptible to new infections with an increased risk of death. Thus, we principally concentrated on the immune-related genes (IRGs) and regarded them as the key molecular events involved in sepsis. After taking the intersection across different platforms, 608 IRGs are screened as potential biomarkers for further analysis. 

\begin{figure*}[h]
	\centering
	\includegraphics[width=0.5\textwidth]{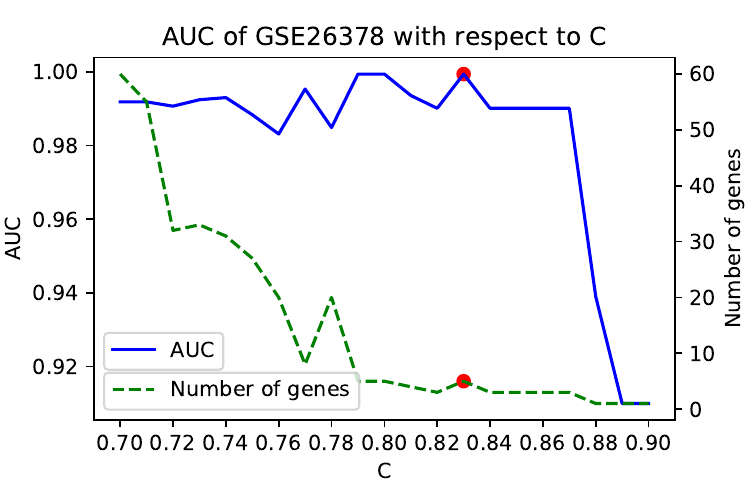}
	\caption{The solid blue line is the AUC with respect to different hyperparameter C between 0.7 and 0.99 with an interval of 0.01. The dash green line is the number of genes with respect to different hyperparameter C between 0.7 and 0.99 with an interval of 0.01. The red dot represents the optimal model with the corresponding $C=0.83$, $gene\_num=5$ and $AUC =0.9994$.}
	\label{fig:2}
\end{figure*}

The recurrent logistic regression (RLR) was then applied on the discovery cohort GSE57065. Different hyperparameter results in multiple gene panels. To select the best gene panel and the hyperparameter, we tried a series of thresholds and evaluated their performance on the independent discovery cohort \uppercase\expandafter{\romannumeral2}, GSE26378. Figure \ref{fig:2} displays the AUROC of these gene panels and indicates that generally a larger C results in a smaller model size N during the optimization. We finalized the model with the highest AUROC up to 0.9951 when C equals to 0.83. The discriminative function of the diagnostic model is 
\begin{equation}
\begin{aligned}
y(x)=&[1+\exp(-0.9305g_{LRRN3}-0.9692g_{IL2RB}\\
&-0.7378 g_{FCER1A}+0.8460 g_{TLR5}\\
&+0.8905g_{S100A12}-0.0153)]^{-1}
\end{aligned}
\end{equation}
which contains five genes, LRRN3, IL2RB, FCER1A, TLR5, and S100A12. We abbreviated the biomarkers by LIFTS, which is composed by the initial letters of each gene. The Genome characteristics of the five genes are listed in Table \ref{tab:2}.

\subsection{The diagnostic capability}
Since the value of logistic model is too concentrated to illustrate, i.e., ranging from 0 to 1, we used the corresponsive part in logits of our diagnostic model to represent the diagnostic ability of each gene and LIFTS. Specifically, the logit is
\begin{equation}
\begin{aligned}
logit=&-0.9305 g_{LRRN3}-0.9692 g_{IL2RB}\\
&-0.7378 g_{FCER1A}+0.8460 g_{TLR5}\\
&+0.8905 g_{S100A12}
\end{aligned}
\end{equation}

The standard difference mean $\bar X-\bar Y$ in effect size between the sepsis and control subjects is displayed in Figure \ref{fig:3}, where $\bar X$ is the mean of logits for the sepsis samples and $\bar Y$ corresponds to normal samples. The five genes individually are qualified to distinguish sepsis samples with the average standard mean difference (SMD) ranging from $1.5$ to $3.5$ and their confidence intervals do not cross zero. Compared with the individual genes, LIFTS achieves a much higher average SMD of $11.6$, suggesting that the weighted gene panel has better diagnostic capability than each of the five genes.

\subsection{Performance comparison across different models}
\begin{figure*}[h]
	\centering
	\includegraphics[width=\textwidth]{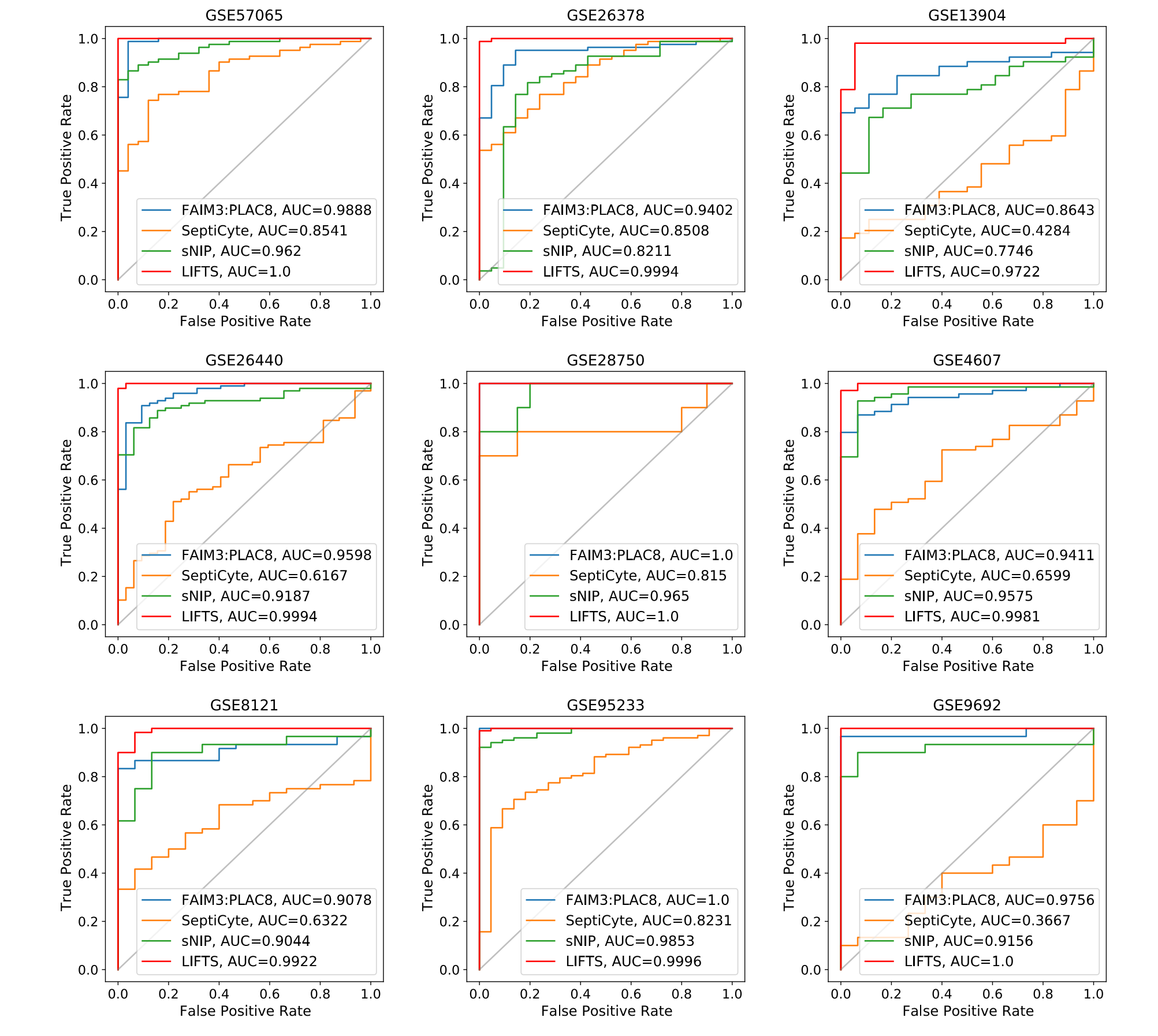}
	\caption{Performance comparison of LIFTS and other existing models in the discovery and validation cohorts. The first two cohorts, GSE57065 and GSE26378, are the discovery cohorts, while the others are validation cohorts.}
	\label{fig:4}
\end{figure*}

LIFTS was evaluated on the nine independent cohorts and compared to existing biomarkers. Figure \ref{fig:4} shows the ROC curves comparison between the LIFTS and three known transcriptome biomarkers, i.e., FAIM3:PLAC8, SeptiCyte Lab, and sNIP. SeptiCyte Lab includes four genes and its risk score is PLAC8/PLA2G7*LAMP1/CEACAM4. sNIP contains three genes and it is represented as (NLRP1-IDNK)/PLAC8. The genes of all these four biomarkers are detectable on the AffyU133P2.

\begin{figure*}[h]
	\centering
	\includegraphics[width=0.6667\textwidth]{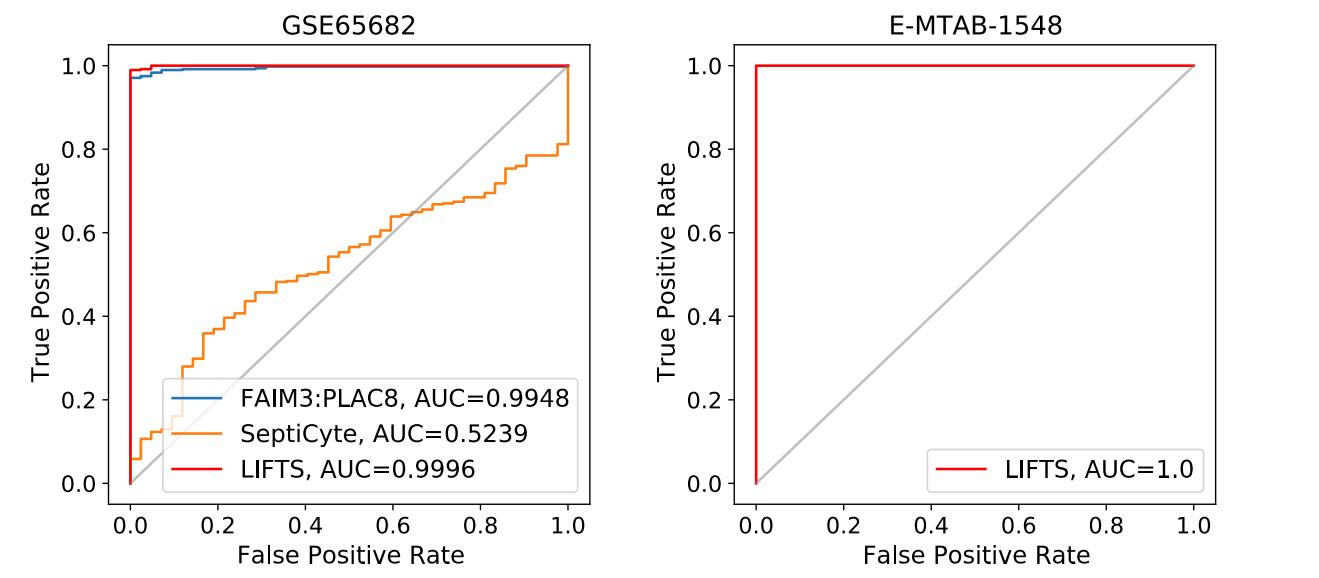}
	\caption{The performance of LIFTS based on two independent cohorts and platforms. NLRP1 does not exist on GSE65682 and PLAC8 is not available on E-MTAB-1548, resulting in the absence of some biomarkers.}
	\label{fig:5}
\end{figure*}

Overall, LIFTS outperforms the other biomarkers on all the validation cohorts except GSE95233. The area under ROC curve (AUC) of LIFTS on each dataset is consistently close to 1. The lowest score, 0.9722 on GSE13904, is still much higher than the other biomarkers. NLRP1 and PLAC8 are not detected on either the AffyU219 or the Agilent platform, so sNIP cannot be applied on dataset GSE65682. Since NLRP1 does not exist on GSE65682 and PLAC8 is not available on E-MTAB-1548, some previous biomarkers could not be evaluated on these two datasets. The AUC of LIFTS on GSE65682 and E-MTAB-1548 are 0.9994 and 1.0, which are superior to its counterparts, indicating the portability of LIFTS among different platforms in diagnostic prediction (Figure \ref{fig:5}).

The AUC curves are related to the standard difference means. Considering LIFTS performance shown in Figure \ref{fig:3} and Figure \ref{fig:4}, the higher AUC value always corresponds to higher standard difference mean. For example, in GSE13904, the AUC value is relatively low and correspondingly the standard difference mean is relatively closer to zero. However, focusing on E-MTAB-1548, the AUC is 1.0 and the standard difference mean is far from zero.

\subsection{Topological and functional analysis of LIFTS}
\begin{figure*}[h]
	\centering
	\includegraphics[width=\textwidth]{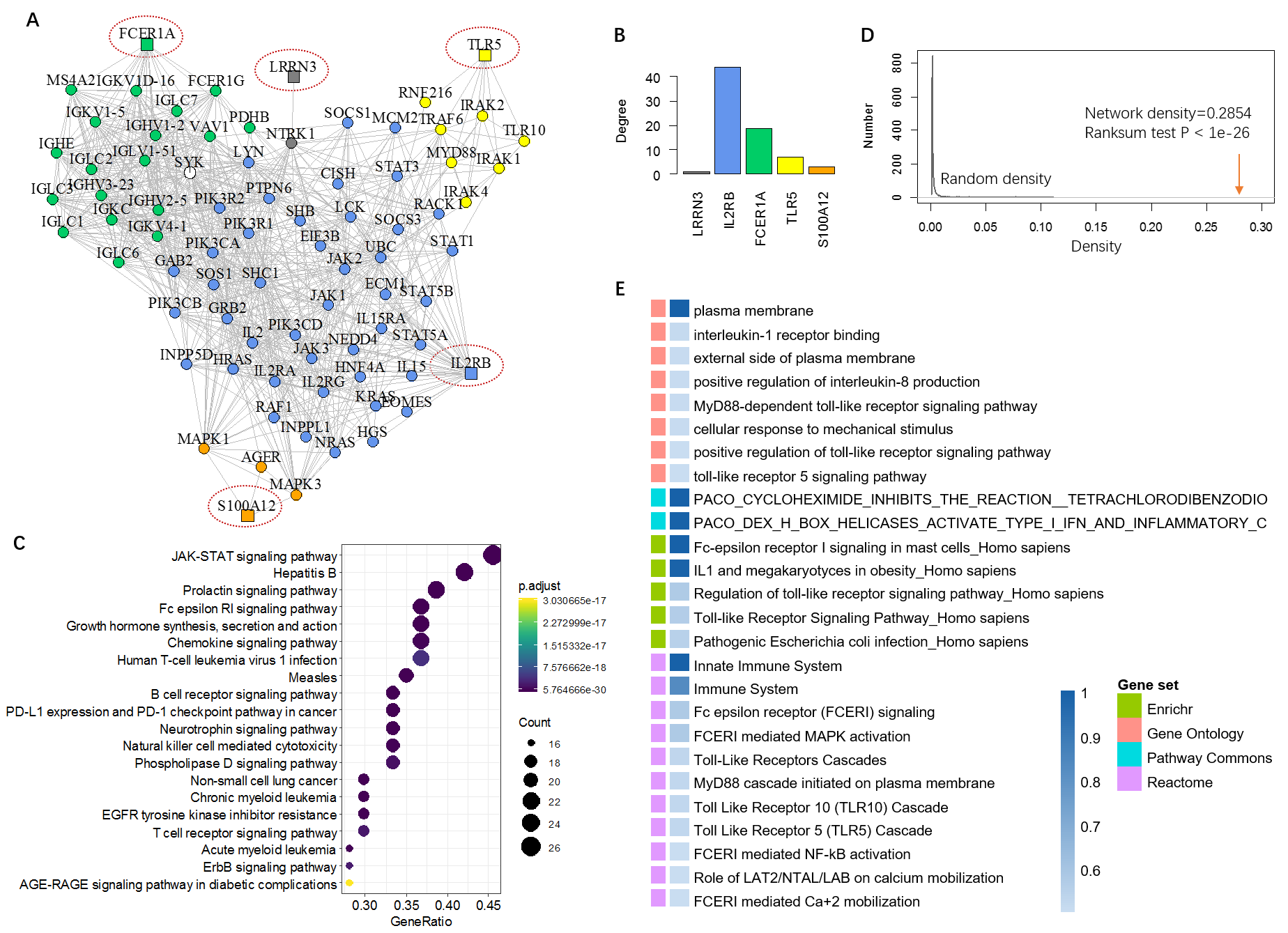}
	\caption{Network and functional analysis of LIFTS. (A) Protein-protein interaction network of LIFTS and their interactors. (B) Interacting degree of LIFTS in the network. (C) Top 20 biological functions overrepresented by genes in the network. (D) Density distribution of simulated networks. (E) Function analysis using the pipeline of network-guided gene set characterization in KnowEnG.}
	\label{fig:6}
\end{figure*}

Proteins usually group together as modules to implement in particular cellular functions through interactions \cite{r20}\cite{r21}\cite{r22}\cite{Cheng_2020}. The interference in protein interactions and new undesired protein interactions can cause diseases \cite{r23}\cite{r24}\cite{r25}.To explore the functions of the genes in LIFTS, we further studied the topological property of the genes physically interacted with the five genes by network analysis (Figure \ref{fig:6}A). Specifically, 1, 45, 19, 7, and 3 partner genes interact with LRRN3, IL2RB, FCER1A, TLR5, and S100A12, respectively (Figure \ref{fig:6}B). These genes are closely connected and involved in specific biological processes, including growth hormone synthesis, secretion and action, chemokine signaling pathway, B cell receptor signaling pathway, T cell receptor signaling pathway, \textit{etc}. (Figure \ref{fig:6}C). For the protein interaction network, the connections among genes are significantly dense than the simulated networks (P $<$ e-26, Rank sum test, Figure \ref{fig:6}D), where we randomly picked up the same number of proteins 10,000 times and calculated their network density distribution. The densities of the simulated networks are mainly less than 0.05 whereas the density of the curated network is 0.2854, indicating the five genes tend to function together with higher network connectivity than other genes.

Given that there are only five genes in LIFTS, standard methods for enrichment analysis may not detect any relevant functional category or pathway. We also employed the network-guided gene set characterization pipeline of KnowEnG \cite{Blatti_2020} for this gene set to better understand their function. Four function resources were used in this pipeline, including Gene Ontology, Enrichr, Pathway Comments, and Reactome. In addition to the functions the partner genes enriched, the five genes are also implemented in pathways of immune system, IL1 and megakaryotyces in obesity, \textit{etc}. Default parameters were used during the analysis.

\section{Discussion}
We developed a novel model to screen the diagnostic biomarkers of sepsis based on the logistic regression. Five genes were identified as a prediction model (LIFTS) with an average AUROC of 0.9959 among 11 cohorts containing in total 1,384 samples. LIFTS demonstrated its robust portability across three different transcriptome platforms, which is much better than its counterparts such as SeptiCyte Lab \cite{r11}. Our analysis thus determined an accurate prediction model and reproducible transcriptome biomarkers that can lay a foundation for clinical diagnostic tests and biological mechanistic studies.

The model was built starting with the immune-related genes. The expression of immune-related genes is response for the dysregulated host immune system to infection in sepsis, so the immune-related genes serve as prior knowledge for the classification model and prevent overfitting, resulting in a robust model for patient heterogeneity. Otherwise, if start with all genes, a different gene panel will be obtained with unexpected noise. The model may get an extremely high performance for the training cohort, but performs worse when it comes to the validation cohorts.

 After filtering the genes, the number of candidate genes was greatly shortlisted, which is also an efficient preprocessing step for feature selection. Some researchers made use of differentially expressed (DE) genes as diagnostic signatures\cite{Liu_2020a}\cite{Scicluna_2017}. However, DE genes are extremely inconsistent among different datasets and platforms. Only a few overlapping DE genes were obtained among the 11 datasets we used (Supplementary Figure S2, S3), leading to obstacles to find a robust classifier based on DE genes.
 
The classical logistic regression can only return the classifier with a given number of genes. Mathematically, our goal is to maximize the performance of classification and minimize the complexity of the diagnostic model simultaneously, requiring a competent algorithm with the ability to filter out irrelevant genes automatically. To this end, an enhanced version of logistic regression, recurrent logistic regression (RLR), was developed using the weight of each term as a measure of the gene importance. Importantly, the selection of features and the construction of classifiers are usually regarded as two independent tasks, but we combined these two tasks together. Thus, the biomarkers are more adaptive to the base model and superior to other models, which use a specific algorithm on previously selected biomarkers. When compared with least absolute shrinkage and selection operator (LASSO), a method that useS L1 regularization to impose sparsity, our results demonstrated that RLR overall outperforms LASSO according to AUROC in the discovery and validation cohorts (Supplementary Figure S1).

During the training process, interestingly, we observed that running logistic regression using different coding languages may lead to different results. In this study, the function LogisticRegression in the sklearn package of python was used. Technically, we applied $L_2$ regularization in our logistic regression processes, which is commonly used in machine learning to reduce model overfitting. In some studies, logistic regression with $L_2$ regularization is called ridge regression \cite{r26}. The advantage of regularization is that it improves numerical stability, not only forces weights to shrink but also copes with the case sophisticatedly when the number of features is larger than the number of samples.

Despite 11 public datasets were taken advantage, all of them were detected using microarray technology. No RNA-seq datasets were included, thereby making our model not applicable for all transcriptome platforms. Therefore, we call for the detection of sepsis using RNA-seq technologies in the near future. Then a large-scale of datasets will be available for further validation, which is able to reduce the risk of the diagnostic model.

Moreover, the proposed method can be used in biomarker identification of other diseases. Since logistic regression is widely used for biomarker identification and several such types of gene expression signatures have been developed for cancers with decent performance, RLR is an upgrade of logistic regression hence it can be applied to the domains where logistic regression can be applied. In terms of the performance, it is superior to logistic regression theoretically, but in practice it depends on a series of factors, such as the detected feature numbers, the sample size, and disease heterogeneity.

In conclusion, the diagnostic biomarkers LIFTS shows higher accuracy and robustness compared to the existing biomarkers when differentiating the sepsis patients from the normal controls. Further clinical trials are needed to confirm the findings in the paper.

\appendices
\section*{Declarations}
\subsection*{Ethics approval and consent to participate}
Not appliable.
\subsection*{Consent for publication}
Not appliable.
\subsection*{Competing interests}
None decleared.
\subsection*{Funding}
This work was supported by the Basic and Applied Basic Research Programs Foundation of Guangdong Province (2019A1515110097).
\subsection*{Authors' contributions}
L.C. conceived the project and wrote the manuscript. Y.Y. and Y.Z. performed research, analyzed data, and drafted the manuscript; S.L., X.Z., M.W., and K.L. supervised the project. All authors read and approved the final manuscript.
\subsection*{Acknowledgements}
Not applicable
\subsection*{Availability of data and materials}
The datasets generated during and/or analysed during the current study are available in GEO database. Source code is available at \href{https://github.com/bio-LIFTS/LIFTS}{https://github.com/bio-LIFTS/LIFTS}.


%\bibliographystyle{IEEEtran}

\bibliography{references}{}

\bibliographystyle{IEEEtran}

\begin{IEEEbiography}[{\includegraphics[width=1in,height=1.25in,clip,keepaspectratio]{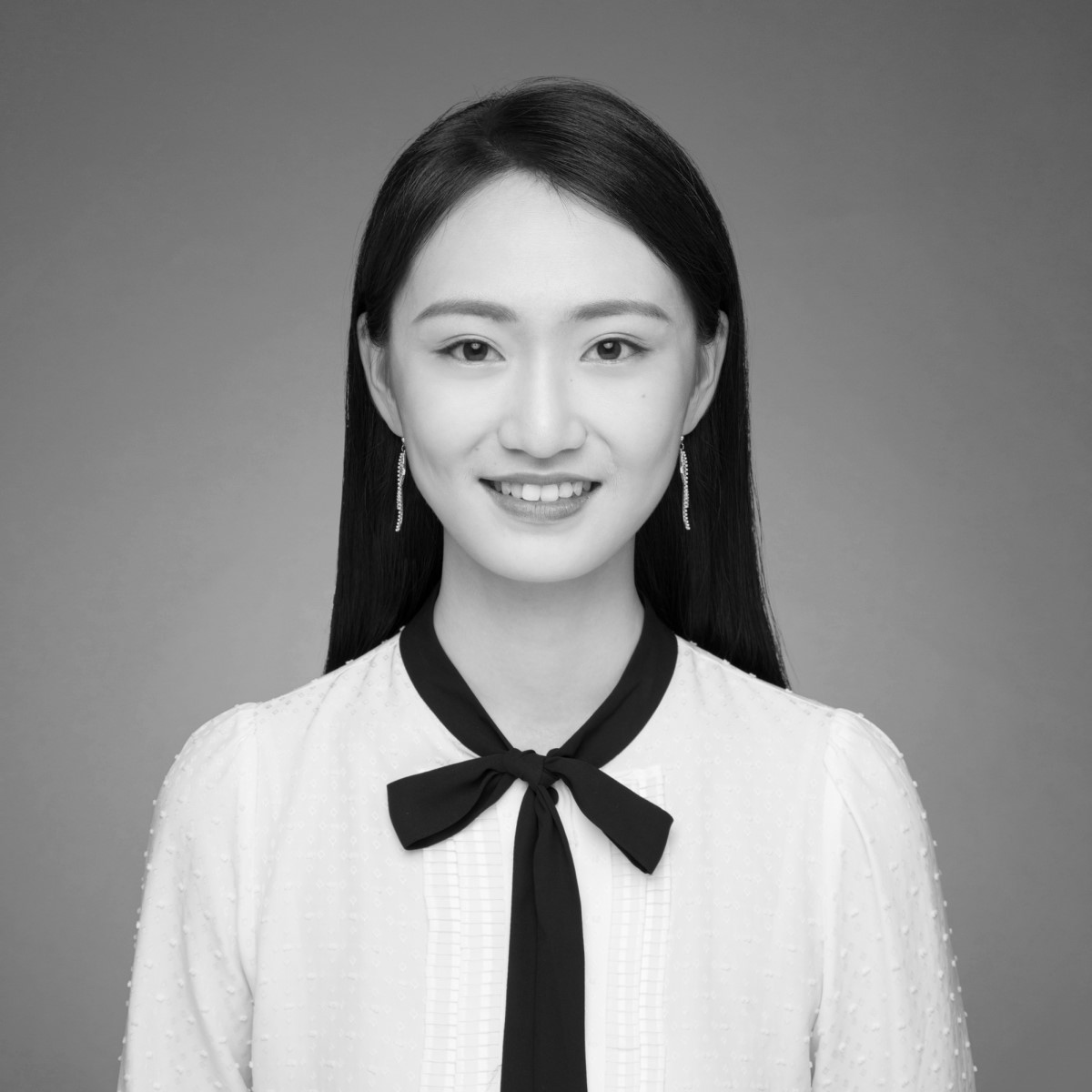}}]{Yueran Yang}
received her BSc. in Mathematics and Applied Mathematics from Shanghai Jiao Tong University. She is currently pursuing her Master's degree at Cornell University. Her research interests include data mining, mathematical modelling and online learning.
\end{IEEEbiography}

\begin{IEEEbiography}[{\includegraphics[width=1in,height=1.25in,clip,keepaspectratio]{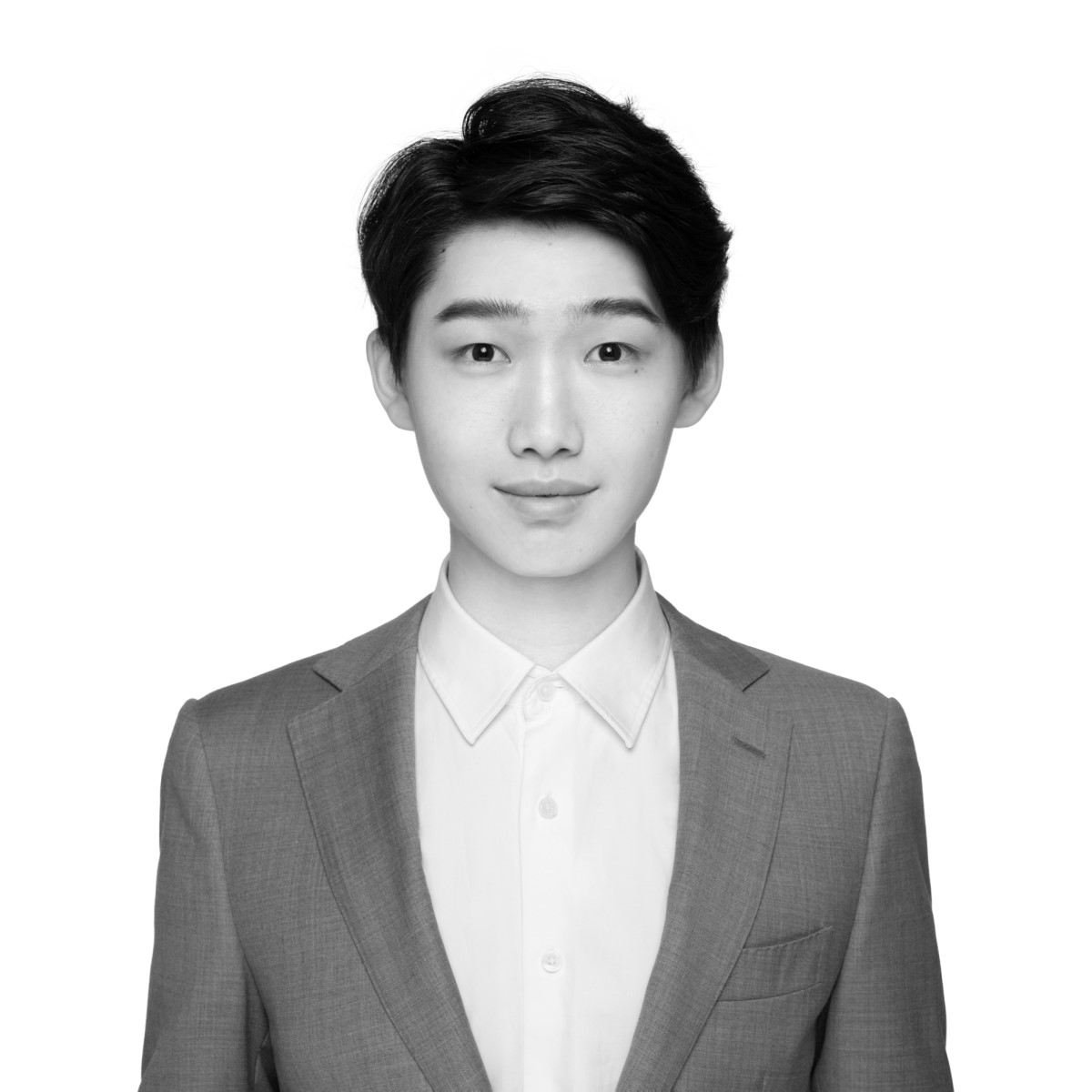}}]{Yu Zhang}
received his BSc. in Mathematics and Applied Mathematics from Shanghai Jiao Tong University. He is currently pursing his Master's degree at Nanyang Technological University. His research interests include data mining, 3D computer vision and natural language processing.
\end{IEEEbiography}

\begin{IEEEbiography}[{\includegraphics[width=1in,height=1.25in,clip,keepaspectratio]{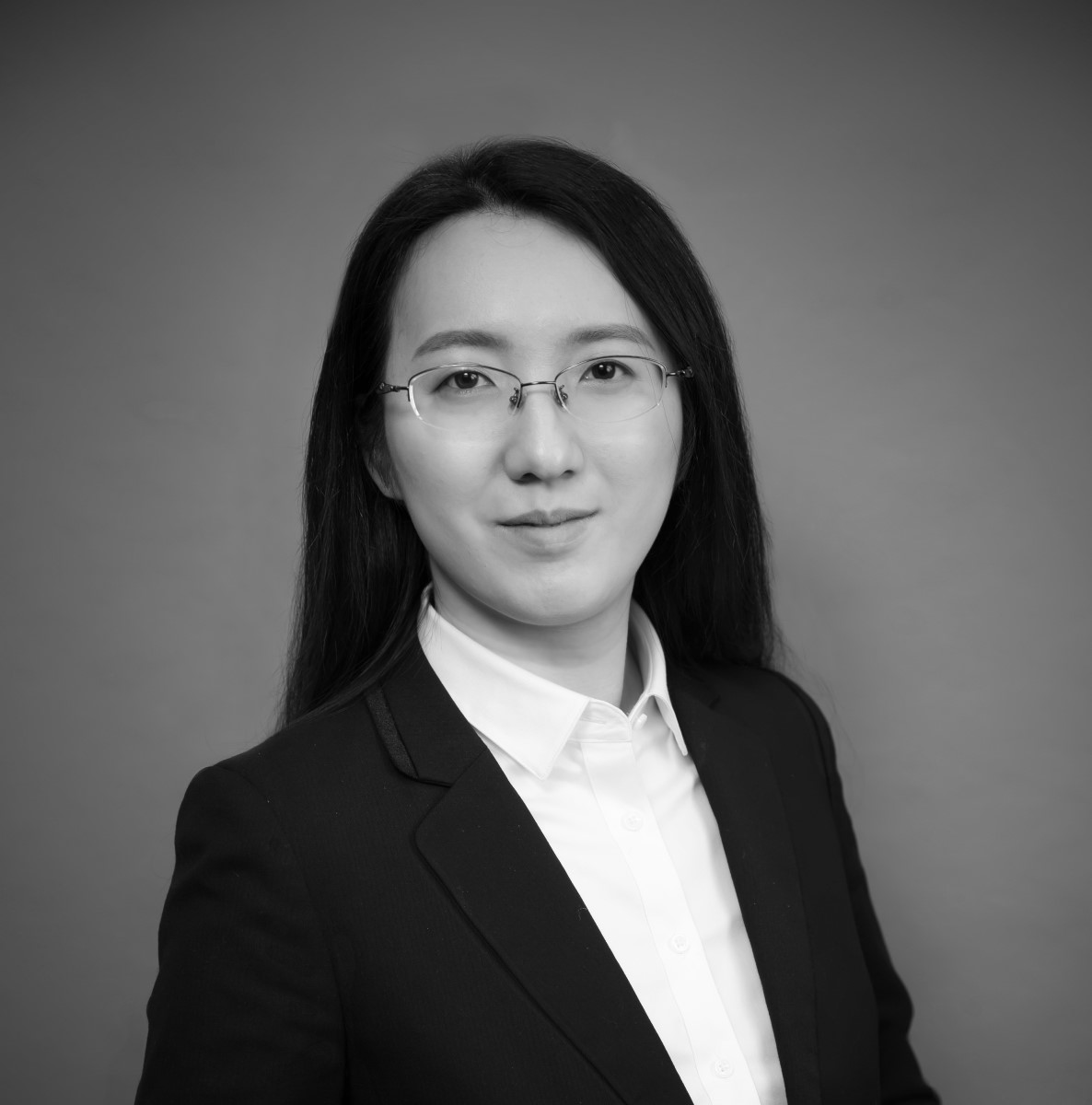}}]{Shuai Li}
is currently a tenure-track assistant professor at John Hopcroft Center of Shanghai Jiao Tong University. She received Ph.D. in computer science and engineering from the Chinese University of Hong Kong. Before that, she received a bachelor's degree in Mathematics from Zhejiang University and a master's degree in Mathematics from the University of the Chinese Academy of Sciences. She has published many top conference papers on ICML/NeurIPS/AAAI/KDD/IJCAI/etc. and serves as reviewers on these conferences. She has visited/interned at many top universities and research labs like UC Berkeley/ University of Alberta/Microsoft/Huawei/Adobe/DeepMind/Tencent AI Lab/etc. She is one of the recipients of Google Ph.D. Fellowship in 2018.
\end{IEEEbiography}

\begin{IEEEbiography}[{\includegraphics[width=1in,height=1.25in,clip,keepaspectratio]{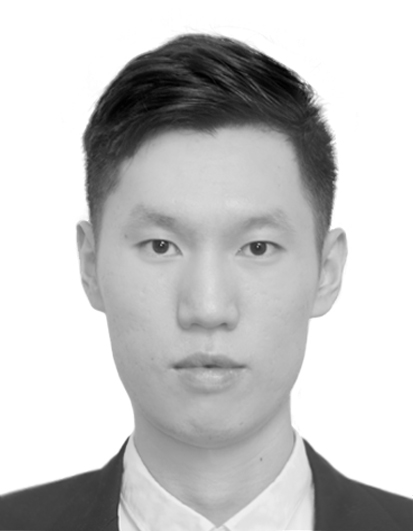}}]{Xubin Zheng}
received the BS degrees from Zhejiang University, China, in 2014 and MS degrees from Hong Kong University in 2016. Currently he is
working toward the doctoral degree in the department of computer science and engineering, the Chinese University of Hong Kong. His research interests include bioinformatics, artificial intelligence, and data mining.
\end{IEEEbiography}

\begin{IEEEbiography}[{\includegraphics[width=1in,height=1.25in,clip,keepaspectratio]{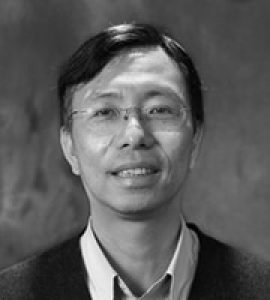}}]{Man-Hon Wong}
received his BS and MP degrees from The Chinese University of Hong Kong in 1987 and 1989 respectively. He got the Ph.D. degree in University of California at Santa Barbara in 1993. Currently he is an associate professor at the department of computer science and engineering, The Chinese University of Hong Kong. His research interests include transaction management, mobile Databases, data replication, distributed systems, expert systems and applications of fuzzy logic.
\end{IEEEbiography}

\begin{IEEEbiography}[{\includegraphics[width=1in,height=1.25in,clip,keepaspectratio]{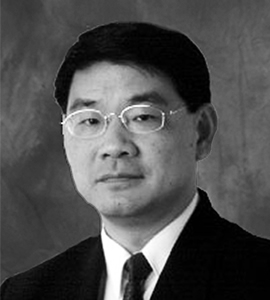}}]{Kwong-Sak Leung}
received his B.Sc. and Ph.D. degrees from the University of London in 1977 and 1980 respectively. He is currently Research Professor and appointed as Professor in the CUHK-BGI Innovation Institution of Trans-omics in the Chinese University of Hong Kong. His research interests include bioinformatics, artificial intelligence, and data mining.
\end{IEEEbiography}

\begin{IEEEbiography}[{\includegraphics[width=1in,height=1.25in,clip,keepaspectratio]{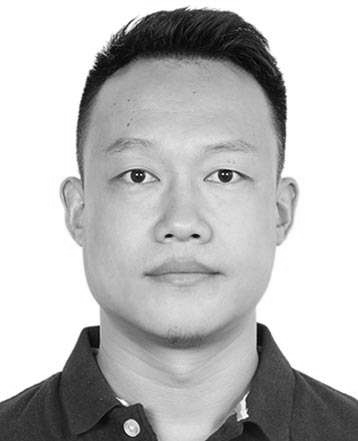}}]{Lixin Cheng}
received BS and MS degrees from Harbin Medical University, China, in 2008 and 2011, and PhD degree from Department of Computer Science \& Engineering at the Chinese University of Hong Kong in 2018. Currently he is working as a PI at Shenzhen People's Hospital, First Affiliated Hospital of Southern University of Science and Technology, China. His research interests include bioinformatics, computational biology, and machine learning.
\end{IEEEbiography}
\enlargethispage{-6in}


%\vfill


\begin{table*}[h]
	\caption{\textbf{Summary of the gene expression cohorts used in this study. }}
	\label{tab:1}
	\centering
	\begin{tabular}{lllllll}
		\toprule
		Series         & Gene Number   & Normal  & Sepsis & Cell type        & Age      & Platform                                                \\
		\midrule
		\multicolumn{7}{l}{Discovery Cohort \uppercase\expandafter{\romannumeral1} }                                                                                                    \\
		GSE57065       & 23521         & 25      & 82     & Whole blood      & Adult    & \multirow{11}{*}{Affymetrix Human Genome U133 Plus 2.0} \\
		\multicolumn{3}{l}{Discovery Cohort \uppercase\expandafter{\romannumeral2} }   &        &                  &          &                                                         \\
		GSE26378       & 23521         & 21      & 82     & Whole blood      & Children &                                                         \\
		\multicolumn{3}{l}{Validation Cohorts \uppercase\expandafter{\romannumeral1} } &        &                  &          &                                                         \\
		GSE95233       & 23521         & 22      & 102    & Whole blood      & Adult    &                                                         \\
		GSE28750       & 23521         & 20      & 10     & Whole blood      & Adult    &                                                         \\
		GSE8121        & 23521         & 15      & 60     & Whole blood      & Children &                                                         \\
		GSE13904       & 23521         & 18      & 52     & Whole blood      & Children &                                                         \\
		GSE26440       & 23521         & 32      & 98     & Whole blood      & Children &                                                         \\
		GSE9692        & 23521         & 15      & 30     & Whole blood      & Children &                                                         \\
		GSE4607        & 23521         & 15      & 69     & Whole blood      & Children &                                                         \\
		\multicolumn{7}{l}{Validation Cohorts \uppercase\expandafter{\romannumeral2} }                                                                                                  \\
		GSE65682       & 19040         & 42      & 479    & Whole blood      & Adult    & Affymetrix Human Genome U219 Array                      \\
		E-MTAB-1548    & 17028         & 15      & 80     & Peripheral blood & Adult    & Agilent Human Gene   Expression 4x44K v2 Microarray \\
		\bottomrule
	\end{tabular}
\end{table*}

\begin{table*}[h]
	\caption{\textbf{Genome characteristics of the genes in LIFTS.}}
	\label{tab:2}
	\centering{}
	\begin{tabular}{lp{6cm}llll}
		\toprule
		Gene symbol & Gene name                                                              & Alignments                   & Chromosomal Location & Degree \\
		\midrule
		LRRN3       & leucine rich repeat neuronal 3                                         & chr7:110731149-110765507 (+) & chr7q31.1            & 1      \\
		IL2RB       & interleukin 2 receptor, beta                                           & chr22:37521886-37545962 (-)  & chr22q13.1           & 45     \\
		FCER1A      & Fc fragment of IgE, high affinity I,   receptor for; alpha polypeptide & chr1:159272125-159277991 (+) & chr1q23              & 19     \\
		TLR5        & toll-like receptor 5                                                   & chr1:223283646-223316624 (-) & chr1q41-q42          & 7      \\
		S100A12     & S100 calcium binding protein A12                                       & chr1:153346183-153348075 (-) & chr1q21              & 3 \\
		\bottomrule   
	\end{tabular}
\end{table*}

\end{document}